\documentstyle[aps,prl]{revtex}

\begin{document}
\draft
\title{Coqblin-Schrieffer impurity in a singular metal}
\author{Yupeng Wang}
\address{Department of Physics, Florida State University, Tallahassee, 
FL 32306, USA, \\
and Institute of Physics, Chinese Academy of Sciences, 
Beijing 100080, \\ 
People's Republic of China}
\author{P. Schlottmann}
\address{Department of Physics, Florida State University, Tallahassee, 
FL 32306, USA}
\maketitle
\begin{abstract}
A Coqblin-Schrieffer impurity of spin $S$ coupled to the boundary of 
an open SU(N)-invariant $t-J$ chain with $N = 2S+2$ is studied. The 
model is integrable as a function of one coupling parameter $v$ for 
arbitrary spin and band filling. The system is coupled to a reservoir 
of electrons and as a function of the chemical potential the host
chain has a critical point corresponding to the van Hove singularity
of the empty band of charges. In the neighborhood of this critical
point we study the low temperature properties of the impurity as a
function of $v$. For constant chemical potential we obtain two 
critical coupling values $v_1=-1/2$ and $v_2=0$ separating phases
in which the impurity spin is completely compensated (Kondo effect), 
partially screened (undercompensated impurity) or decoupled from the 
chain (free spin). The partial compensation occurs as a consequence 
of a thermally {\it activated Kondo effect} in the region $v_1 < v 
< v_2$. The possible realization of such a {\it finite temperature 
Kondo effect} in narrow-gap semiconductors is discussed.
\end{abstract}

The unusual non-Fermi-liquid properties observed in several heavy-fermion 
compounds \cite{1} are frequently attributed to the existence of a 
quantum critical point (QCP). For instance, Kondo impurities in a 
multi-channel electron host \cite{2} provide such a QCP in the 
overscreened case with non-Fermi-liquid behavior in a variety of 
low-temperature thermodynamic quantities \cite{3}. Magnetic impurities 
coupled to a Luttinger liquid may behave rather differently from 
those in a Fermi liquid \cite{4}. Many impurity QCP have been
predicted in exactly solvable one-dimensional correlated systems 
\cite{5,6}. Renormalization group studies indicate that quantum impurity 
models flow toward a low-temperature fixed point corresponding to 
conformally invariant boundary conditions \cite{7}. Also Wilson's
numerical renormalization group maps the three-dimensional Kondo model 
to a one-dimensional Fermi lattice with a boundary impurity \cite{8}.
An impurity QCP is then given by the parafermionic sector of a 
Wess-Zumino-Novikov-Witten model. 

Another topic of recent interest is the Kondo screening in Fermi systems 
with pseudo-gap, i.e., in systems where the density of states 
$\rho(\epsilon)$ has a power-law dependence on the energy $\epsilon$ 
near the Fermi level, $\rho(\epsilon) \sim \epsilon^r$. Using a 
renormalization-group (RG) analysis, Withoff and Fradkin \cite{9} 
showed that there could be a critical value $J_c$ for the Kondo 
coupling constant, so that for $J>J_c$ Kondo screening occurs at 
low temperatures, while for $J<J_c$ the impurity decouples from the 
host and its magnetic susceptibility is divergent at $T=0$. Here 
the host itself is not a regular Fermi liquid and its low-temperature 
specific heat and susceptibility behave as $C \sim T^{r+1}$ and 
$\chi\sim T^r$, respectively. As discussed recently \cite{10}, 
the critical behavior of the impurity strongly depends on the host 
properties and appears to be non-universal. Due to the crossover
regime from Kondo compensated to unscreened impurity spin almost
all perturbation techniques fail in the critical regime. Under 
these circumstances, exact results, even for a special situation, 
may shed light onto this complicated subject.

In this letter, we study the Coqblin-Schrieffer impurity model in a 
correlated singular metal with $\rho(\epsilon)\sim \epsilon^{-1/2}$. 
In this case, the density of states at the Fermi level is divergent 
rather than zero. Such a situation can be realized when the Fermi 
level falls at the one-dimensional van Hove singularity of the electron
band. The Hamiltonian is given by
\begin{eqnarray}
H &=& - \sum_{j=1}^{L-1} \sum_{s=-S}^S {\cal P} \Bigl(c_{j,s}^\dagger 
c_{j+1,s} + h.c. \bigr) {\cal P} + \sum_{j=1}^{L-1} \sum_{s,s'=-S}^S
\Bigl( c_{j,s}^\dagger c_{j,s'} c_{j+1,s'}^\dagger c_{j+1,s} +
n_jn_{j+1} \Bigr) \nonumber \\
&+& J \sum_{s,s'=-S}^S c_{1,s}^\dagger c_{1,s'} f_{s'}^\dagger f_{s}
+ Vn_1 - \mu \sum_{j=1}^L n_j \ , 
\end{eqnarray}
where $c_{j,s}^\dagger$ ($c_{j,s}$) is the creation (annihilation) 
operator of an electron at the site $j$ with spin component $s$, 
$-S \le s \le S$, $n_j \equiv \sum_s n_{j,s}=\sum_s c_{j,s}^\dagger 
c_{j,s}$ is the particle number operator, and $\cal{P} \cdots \cal{P}$ 
excludes the multiple occupation of each site $n_j \leq 1$. Here $J$ 
is the Kondo coupling constant, $V$ is a boundary potential, $\mu$ 
is the chemical potential, $f_s^\dagger$ ($f_s$) is the creation 
(annihilation) operator of the impurity with $\sum_s f_s^\dagger 
f_s=1$, so that there is always exactly one electron localized at 
the impurity. The Hamiltonian (1) describes a one-dimensional 
SU($N$)-invariant $t-J$ model ($N=2S+2$) with a boundary impurity. 
Related $t-J$ models embedding a Coqblin-Schrieffer or Anderson 
impurity have been studied in Ref. \cite{schlot}. 

We consider model (1) coupled to a reservoir of electrons, 
constituted by bands without direct interaction with the 
impurity. In this way the number of carriers in the chain changes
with temperature if the chemical potential is fixed. At $\mu_c = 
-2$ the ground state of the $t-J$ model undergoes an insulator-metal 
transition from insulating for $\mu < \mu_c$ to a low carrier density 
conductor for $\mu > \mu_c$. At this critical point, the chain is a 
singular metal in the sense that the density of states for low-lying 
excitations diverges as $\rho(\epsilon)\sim\epsilon^{-1/2}$, which 
is due to the van Hove singularity. 

The Hamiltonian (1) is exactly solvable via Bethe's 
{\it Ansatz} as a function of one parameter $v$ determining $J$ and
$V$ via $J=(1-v^2)^{-1}$, $V=1 + v/(1-v^2)$. For this parametrization
the reflection matrix $R(q)$ of the electrons off the impurity 
satisfies the reflection Yang-Baxter equation (YBE)
\begin{eqnarray}
S(q_i-q_j)R(q_i)S(q_i+q_j)R(q_j)=R(q_j)S(q_i+q_j)R(q_i)S(q_i-q_j),
\end{eqnarray}
where $S(q_i\pm q_j)$ denotes the two-body scattering matrix between
electrons in the bulk \cite{Suther}. Note that $v$ is required to be 
real to ensure a Hermitian Hamiltonian. Following the standard method 
outlined by Sklyanin \cite{13} and its generalizations \cite{14} we 
obtain the Bethe ansatz equations (BAE) diagonalizing the model in
terms of $2S+1$ sets of rapidities $\{q_j^{(r)}\}$, $r=0,\cdots,2S$, 
\begin{eqnarray}
&&\left( \frac{q_j^{(0)} - \frac i2}{q_j^{(0)} + \frac i2} \right)^{2L+1}
= - \frac{q_j^{(0)} - i(\frac{3}{2}-v)}{q_j^{(0)} + i(\frac{3}{2}-v)} 
\prod_{p=\pm} \prod_{l\neq j}^{M_0} \frac{q_j^{(0)} - pq_l^{(0)} - 
i}{q_j^{(0)} - pq_l^{(0)} + i} \prod_{\alpha=1}^{M_1} \frac{q_j^{(0)} - 
pq_\alpha^{(1)} + \frac i2}{q_j^{(0)} - pq_\alpha^{(1)} - \frac i2} 
, \nonumber \\
&&\prod_{p=\pm} \prod_{t=0,2} \prod_{j=1}^{M_t} \frac{q_\alpha^{(1)} 
- pq_j^{(t)} - \frac i2}{q_\alpha^{(1)} - pq_j^{(t)} + \frac i2} = 
\frac{q_\alpha^{(1)} + iv}{q_\alpha^{(1)} - iv} \ \frac{q_\alpha^{(1)} 
+ i(1-v)}{q_\alpha^{(1)} -i (1-v)} \prod_{p=\pm} \prod_{\beta\neq\alpha} 
\frac{q_\alpha^{(1)} - pq_\beta^{(1)} - i}{q_\alpha^{(1)} - 
pq_\beta^{(1)} + i} , \\
&&\prod_{p=\pm} \prod_{t=r\pm1} \prod_{l=1}^{M_t} \frac{q_j^{(r)} -
pq_l^{(t)} - \frac i2}{q_j^{(r)} - pq_l^{(t)} + \frac i2} = 
\prod_{p=\pm} \prod_{m\neq j}^{M_r} \frac{q_j^{(r)} - pq_m^{(r)} - 
i}{q_j^{(r)} - pq_m^{(r)} + i} , {~~~~}r>1 \nonumber
\end{eqnarray}
where $M_r=\sum_{l=r}^{2S}n_{l-S}$ ($r=0,\cdots, 2S$) and $M_{2S+1} 
= 0$. The impurity gives rise to the factors involving $v$ in the 
first and second set of equations (3). The set of rapidities 
$\{q_j^{(0)}\}$ refers to the charges, while the remaining sets 
for $r = 1, \cdots ,2S$ correspond to the spin degrees of freedom. 
The energy of the system is given by
\begin{eqnarray}
E=\sum_{j=1}^{M_0} \left[2 - \frac 1{\bigl(q_j^{(0)}\bigr)^2 + 
\frac14} - \mu \right].
\end{eqnarray}

We first briefly summarize the properties of the chain without the 
impurity. In the thermodynamic limit the solutions of the BAE are 
given in terms of string excitations of arbitrary length $n$ of the 
form $q^{(r)} = \Lambda + i(n - 1 - 2l)/2$ with $l = 0, \cdots, n-1$ 
for each of the sets of rapidities. Here $\Lambda$ represents the
rapidity of the center of mass of the string. We denote with 
$\varphi_n^{(r)}(\Lambda) = T \ln[\eta_{r,n}(\Lambda)]$ the dressed 
energy of the corresponding string excitations, which satisfy the
following integral equations obtained by minimizing the free 
energy \cite{15,16,17}  
\begin{eqnarray}
\ln\bigl(1 + \eta_{r,n}(\Lambda) \bigr) &=& \frac{\epsilon_{r,n}(\Lambda)}T 
+ \sum_{m=1}^\infty A_{nm} \star \ln(1+\eta_{r,m}^{-1}) \nonumber\\
&-& \sum_{m=1}^\infty B_{nm} \star \bigl[ \ln\bigl( 1+\eta_{r-1,m}^{-1} 
\bigr) + \ln\bigl(1+\eta_{r+1,m}^{-1}\bigr) \bigr] \ ,
\end{eqnarray}
where $T$ is the temperature and $\star$ represents a convolution.
Here $\epsilon_{0,n}(\Lambda) = n(2-Sh-\mu) - 2\pi a_n(\Lambda)$, 
$\epsilon_{r,n}=nh$ for $r=1,\cdots,2S$, and $h$ is the magnetic field. 
The integration kernels are $A_{nm} = a_{m+n} + a_{|m-n|} + 
2\sum_{l=1}^{min(m,n)-1} a_{m+n-2l}$ and $B_{mn} = \sum_{l=1}^{min(m,n)} 
a_{m+n-2l+1}$, where $a_n(\Lambda) = |n|/[2\pi(\Lambda^2+n^2/4)]$. 
In Eqs. (5) we assumed that $\eta_{2S+1,n} = \eta_{-1,n} = \infty$ 
for all $n$. The free energy of the host is given by $F = L f_{host}$, 
where
\begin{equation}
f_{host} = - T \sum_n \int d\Lambda a_n(\Lambda) \ln[1 + 
\eta_{0,n}^{-1}(\Lambda)] \ .
\end{equation}

In the ground state all the rapidities are real, i.e. only states with 
$n=1$ can be occupied. From the expression of the energy it is seen that 
in zero-field for $\mu \le \mu_c = -2$ the band of charge rapidities 
($r = 0$) is empty and the system is an insulator. For $\mu$ slightly 
larger than $\mu_c$ the system has a low-density of carriers. The dressed 
energy for the charges for this case is given by
\begin{equation}
\varphi_1^{(0)}(\Lambda) = 2-Sh-\mu - 2\pi a_1(\Lambda) \approx -Sh - \mu
- 2 + 16 \Lambda^2 \ , 
\end{equation}
where in the last step we assumed $|\Lambda|$ small. The Fermi points are
determined by $\varphi_1^{(0)}(\pm Q) = 0$, i.e. $Q = \frac{1}{4} 
\sqrt{2+\mu+Sh}$, which in zero-field tends to zero at $\mu_c$. The 
ground state energy and susceptibility for small $Q$ are then given by 
\begin{equation}
E_{host}/L = - \frac{2}{3\pi} \bigl(Sh + \mu + 2\bigr)^{3/2} \ \ , \ \ 
\chi_{host}/L = (S^2 / 2\pi) \bigl(Sh + \mu + 2\bigr)^{-1/2} \ .
\end{equation}
The expression for $\chi_{host}$ diverges with a square root singularity
as a consequence of the one-dimensional van Hove singularity of the
empty charge rapidity band.

Similarly, the low-temperature thermodynamics of the bulk of the chain 
can be derived to leading order. The main contribution to the free
energy arises from $\varphi_1^{(0)}$; substituting Eq. (7) into Eq. (6), 
we obtain
\begin{eqnarray}
f_{host} = - c_1 T^{\frac32} - c_2 T^{-\frac12} h^2 + \cdots \ ,
\end{eqnarray}
where $c_1 = \int \ln(1+e^{-x^2}) dx$ and $c_2 = \frac18 S^2 \int 
\cosh^{-2}(x^2/2) dx$. The specific heat and the susceptibility are then
\begin{eqnarray}
C_{host}/L = \frac{3}{4} c_1 T^{\frac12} + \cdots \ \ , \ \
\chi_{host}/L = 2 c_2 T^{-\frac12} + \cdots \ .
\end{eqnarray}
This result indicates that a non-interacting charge model already gives 
exactly the leading order of the low-temperature thermodynamics. Because 
of the low carrier density the interactions between particles can 
plausibly be omitted, since they contribute only to subleading order.

In general the impurity can give rise to imaginary modes, which 
correspond to boundary bound states involving impurity states. From 
the first set of the BAE, Eqs. (3), we see that $q^{(0)}=i(3/2-v)$ for 
$v<3/2$ is a solution in the thermodynamic limit $L\to\infty$. The
energy associated with this bound state is 
\begin{equation}
\epsilon_b = 2 - \mu + {1 \over {(v-1)(v-2)}} \ , 
\end{equation}
so that it is stable only in a small interval of $v$ for $v > 1$. This
corresponds to ferromagnetic coupling of the impurity to the chain. 
In the antiferromagnetic Kondo coupling regime $J>0$ ($|v|<1$), the 
energy of this state is a high energy excited charge bound state. 

A spin-charge bound state should exist, because when $v \to -1^+$
then $J \to +\infty $ and $V \to -\infty$, i.e. both the Kondo coupling
and the attractive boundary potential enforce the formation of a 
stable boundary bound state between the impurity and one bulk particle. 
A detailed analysis of the second set of BAE shows that $q^{(1)}=iv$ 
is a solution if $q^{(0)}=i(v+1/2)$ is a solution of the first set of
Eqs. (3). This is indeed the case in the interval $-1<v<-1/2$, 
where this state describes the spin singlet pairing of the impurity 
with one electron of the bulk. Interestingly, this bound state is 
stable even in the insulating phase ($\mu < \mu_c$), as long as $\mu
\ge 2 + 1/(v+v^2)$, i.e., the Kondo screening exist if the Kondo 
coupling is strong enough to overcome the energy gap of the charges.

Below we examine how the impurity behaves in the neighborhood of the
quantum critical point $\mu_c = -2$. Two situations have to be 
distinguished in the ground state, namely, for $-1<v<-1/2$, the 
local moment of the impurity is quenched into a singlet by one host 
electron, while for $-1/2 \leq v < 1$, the ground state is 
$(2S+1)$-fold degenerate and an arbitrary small magnetic field 
induces a finite magnetization $S$. In other words, there is a 
first order phase transition at $v = -1/2$. Such a critical point 
is indeed very similar to the $J_c$ found in Ref. [9]. 

The free energy of the impurity is driven by the host via the 
functions $\eta_{r,n}(\Lambda)$, 
\begin{eqnarray}
f_{imp} = - T \sum_{r=0,1} \sum_n \int d\Lambda \phi_{r,n}(\Lambda) 
\ln\bigl[1 + \eta_{r,n}^{-1}(\Lambda) \bigr] \ , 
\end{eqnarray}
with
\begin{eqnarray}
\phi_{0,n}(\Lambda) &=& \frac12 \sum_{l=1}^n a_{|n-2l+2v-2|}(\Lambda) 
sign(n-2l+2v-2) \ , \nonumber \\
\phi_{1,n}(\Lambda) &=& \frac12 \sum_{l=1}^n a_{|n-2l+2v+1|}(\Lambda) 
sign(n-2l+2v+1) \nonumber \\
&+& \frac12 \sum_{l=1}^n a_{|n-2l-2v+3|}(\Lambda) sign(n-2l-2v+3)  
\end{eqnarray}
for $-\frac12\leq v<1$, and 
\begin{eqnarray}
\phi_{0,n}(\Lambda) &=& \frac12 \sum_{l=1}^n a_{|n-2l+2v+2|}(\Lambda)
sign(n-2l+2v+2) \nonumber \\
&+& \frac12 \sum_{l=1}^n a_{|n-2l+2v-2|}(\Lambda) sign(n-2l+2v-2) 
\nonumber \\
&+& \frac12 \sum_{l=1}^n a_{|n-2l-2v-2|}(\Lambda) sign(n-2l-2v-2) 
\nonumber \\
\phi_{1,n}(\Lambda) &=& 0 
\end{eqnarray}
for $-1<v<-\frac12$. Here we have omitted the unstable charge boundary 
bound state and the unstable charge-spin bound state for $-1/2 \leq v 
< 1$, as well as the excitations arising from them, since their 
contributions are exponentially small at sufficiently low temperatures.
In Eq. (12) we also omitted the energy of the stable bound state for
$-1<v<-\frac12$, which is just an additive temperature and field 
independent constant, $\epsilon_{bs} = 2 -\mu + 1/[v(v+1)]$.

In the ground state only the $n = 1$ strings are relevant. Using 
Eq. (7) and following arguments similar to those used to derive 
Eq. (8), we obtain
\begin{eqnarray}
\chi_{imp} &=& S \delta(h) - \frac{S^{3/2}}{4\pi} \frac{1}{(3-2v)} 
h^{-\frac12} + \cdots,{~~~}{\rm for}{~~~} - \frac12\leq v<1, \nonumber \\
\chi_{imp} &=& - \frac{S^{3/2}}{4\pi} \frac{(4v^2 + 12v - 3)}{(9-4v^2)
(2v+1)} h^{-\frac12} + \cdots, {~~~}{\rm for}{~~}-1<v<-\frac12.
\end{eqnarray}
The second term for $-1/2\leq v<1$ is negative, which means that 
the local moment is not completely free as a consequence of the Kondo 
coupling of the impurity to the bulk modes, which yield contributions 
to subleading order. Note that for $-1<v<-\frac12$ $\chi_{imp}$ is
negative because one spin from the host is locked into the singlet 
boundary bound state. From Eq. (8) we have that the host susceptibility 
is also proportional to $h^{-\frac12}$, so that as expected the 
critical behavior of the impurity is locked into that of the host.

At low temperatures the charge strings for $n > 1$ are gapped and 
for $\mu \approx \mu_c$ we have
\begin{eqnarray}
\eta_{0,n}(\Lambda) \approx \exp\biggl[\frac{n(2 - \mu -Sh) - 2\pi 
a_n(\Lambda)}T \biggr] \ .
\end{eqnarray}
Hence, their contributions to the free energy are exponentially small
and can be neglected, except for $\eta_{0,1}(\Lambda)$ which may have a 
Fermi surface and contributes to leading order to the low-$T$ 
thermodynamics. In addition, at low $T$ and for $\mu \approx \mu_c$
the spin and charge degrees of freedom are only weakly coupled. Hence,
all $\eta_{r,n}(\Lambda)$ for $r \ge 1$ tend to constants, so that the 
integral equations determining the $\eta_{r,n}$ reduce to algebraic ones
and are easily solved \cite{16,18}
\begin{eqnarray}
\eta_{r,n} + 1 = \frac{\sinh[(n+r)x_0] \sinh[(n+2S-r+1)x_0]}{\sinh(rx_0)
\sinh[(2S+1-r)x_0]} \ ,
\end{eqnarray}
where $x_0 = h / 2T$.

For the low-temperature thermodynamics of the impurity, we distinguish 
three cases: (i) $0<v<1$, (ii) $-1/2\leq v\leq 0$, and (iii) $-1<v<-1/2$. 
For case (i) the free energy of the impurity can be approximated by
\begin{eqnarray}
f_{imp} \approx \frac12 T \int d\Lambda a_{3-2v}(\Lambda) 
\ln\Bigl[1+\eta_{0,1}^{-1}(\Lambda)\Bigr] - Sh - T \sum_{n=1}^{\infty}
\ln(1+\eta_{1,n}^{-1}).
\end{eqnarray} 
Using Eq. (17) the last two terms can be rewritten as $-T \ln\{
\sinh[(2S+1)x_0]/\sinh(x_0)\}$, which is just the free energy of the
free spin. This clearly indicates that in the region (i) the charge
and spin degrees of freedom are decoupled. The specific heat of the 
charges is related to that of the bulk by $C_{imp}=[1/(4v-6)] C_{bulk}
\sim T^{\frac 12}$ (as in Eq. (15) this contribution is negative), 
while the spin part gives rise to a Schottky anomaly. The zero-field
residual entropy $S_{res}(T\to0)$ is $\ln(2S+1)$, which exactly 
corresponds to a free spin $S$. The impurity susceptibility follows 
a Curie law with a subleading term $[1/(4v-3)] \chi_{bulk} \sim 
T^{-\frac12}$, which arises from the charges.

For case (ii), $-1/2 \leq v < 0$, again the charge and spin 
degrees of freedom decouple. While the charge part is identical 
to case (i) (and so is the $T^{\frac12}$ dependence of the specific 
heat and the $T^{-\frac12}$ dependence of the susceptibility), the 
spin contribution to the impurity free energy is now $- Sh - T 
\sum_{n=2}^{\infty} \ln(1+\eta_{1,n}^{-1}) = - T \ln\{\sinh[(2S+2)x_0]
/\sinh(2x_0)\}$. Consequently, the Schottky anomaly and the Curie 
constant are changed to that of an effective spin that in general 
is neither integer nor half-integer. The residual zero-field entropy 
at finite but low $T$ is then $S_{res} = \ln(S+1)$, i.e. it is 
reduced by the Kondo coupling with respect to the free spin. Hence, 
as a function of $v$ the entropy has a jump at $v=0$ of
\begin{eqnarray}
\Delta S_{res} = \ln \left[ \frac{2S+1}{S+1} \right] \ .
\end{eqnarray}
The reduction of the entropy indicates that partial Kondo screening 
occurs at low but finite temperatures. However, the degeneracy of 
the ground state in zero-field is exactly ($2S+1$)-fold and the entropy
is continuous as a function of $T$. 

This unusual phenomenon can be understood within the following 
picture. In the ground state there are no excited states to screen 
the impurity and thus there is no Kondo effect. However, at finite 
temperature, there exist bulk excitations which partially screen 
the local moment. It is this {\it activated Kondo effect} that 
induces a discontinuity in the entropy as a function of $v$ at $v=0$. 
The situation encountered here is very different from the one in a 
Fermi liquid or in a Luttinger liquid, where there are sufficient 
conduction electrons to compensate the impurity spin at any 
temperature. Hence, the lower $T$, the stronger the Kondo screening. 
In the present case, however, $T$ itself controls the number of 
carriers and therefore the Kondo compensation. Experimentally,
the {\it activated Kondo effect} could occur in dilute magnetically 
doped narrow-gap semiconductors, where thermally activated 
carriers may induce a finite temperature Kondo effect. For $v>0$, 
the larger boundary potential $V$ dominates over the Kondo coupling 
$J$ and repels the bulk particles from the impurity. This explains 
the absence of a temperature activated Kondo effect in this 
parameter region. 

Finally for $-1<v<-1/2$, case (iii), we have from Eq. (14) that at 
very low temperature the impurity only contributes to the free 
energy in the charge sector because $\phi_{1,n}=0$. We obtain for
this case
\begin{eqnarray}
C_{imp} &=& \frac{(-4v^2 - 12v + 3)}{(9-4v^2)(2v+1)}C_{host} \ ,
\nonumber\\
\chi_{imp} &=& \frac{(-4v^2 - 12v + 3)}{(9-4v^2)(2v+1)}\chi_{host}
\ , \\
S_{res}(0) &=& S_{res}(T \to 0) = 0 \ , \nonumber
\end{eqnarray}
which means the critical behavior of the impurity is locked into 
that of the bulk, as in the case of Fermi liquids.

In conclusion, we have studied the properties of a Coqblin-Schrieffer 
impurity in a singular metal. At low temperatures three situations 
have to be distinguished, namely, the standard Kondo screening of 
the impurity into a singlet state, an ``activated Kondo effect''
in which a partial spin compensation is induced via thermal 
activation, and for repulsive coupling at the boundary the impurity 
spin is asymptotically free. As predicted in Ref. [9] for a semi-metal 
a critical value for the coupling constant is required, below 
which the zero temperature Kondo screening disappears. We suggest
that a possible realization of the thermally ``activated Kondo 
effect'' could be in very-narrow-gap semiconductors. 

We acknowledge the support by the National Science Foundation 
and the Department of Energy under grants No. DMR98-01751 and
No. DE-FG02-98ER45797.  Y. Wang is also supported by the 
National Science Foundation of China.


\begin{references}
\bibitem{1}M.B. Maple et al., J. Low Temp. Phys. {\bf 99}, 223 (1995); 
H. von L\"{o}hneysen et al., Physica (Amsterdam) B {\bf 230-232}, 550 
(1997).
\bibitem{2}P. Nozi\'{e}res and A. Blandin, J. Phys. (Paris) {\bf 41}, 
193 (1980).
\bibitem{3}N. Andrei and C. Destri, Phys. Rev. Lett. {\bf 52}, 364 
(1984); P.B. Wiegmann and A.M. Tsvelik, JETP Lett. {\bf 38}, 596 (1983); 
P. Schlottmann and P.D. Sacramento, Adv. Phys. {\bf 42}, 441 (1993).
\bibitem{4}D.-H. Lee and J. Toner, Phys. Rev. Lett. {\bf 69}, 3378 (1992); 
A. Furusaki and N. Nagaosa, Phys. Rev. Lett. {\bf 72}, 892 (1994); 
P. Fr\"{o}jdh and H. Johannesson, Phys. Rev. Lett. {\bf 75}, 300 (1995).
\bibitem{5}Y. Wang et al., Phys. Rev. Lett. {\bf 79}, 1901 (1997); 
Y. Wang, Phys. Rev. B {\bf 56}, 14045 (1997); J. Dai, Y. Wang and U. 
Eckern, Phys. Rev. B {\bf 60}, 6594 (1999).
\bibitem{6}A. Zvyagin and H. Johannesson, Phys. Rev. Lett. {\bf 81}, 
2751 (1998); A. Zvyagin and P. Schlottmann, J. Phys. A {\bf 31}, 1981 
(1998); Nucl. Phys. B {\bf 565} [FS], 555 (2000).
\bibitem{7}A.W.W. Ludwig and I. Affleck, Nucl. Phys. B {\bf 428} [FS], 545 
(1994); I. Affleck in Springer Series in Solid State Sciences, {\bf 118}, 
ed. by A. Okiji and N. Kawakami (Springer-Verlag, 1994) and references 
therein.
\bibitem{8}K.G. Wilson, Rev. Mod. Phys. {\bf 47}, 773 (1975).
\bibitem{9}D. Withoff and E. Fradkin, Phys. Rev. Lett. {\bf 64}, 1835 
(1990).
\bibitem{10}C.R. Cassanello and E. Fradkin, Phys. Rev. B {\bf 53}, 15079
(1996); {\bf 56}, 11246 (1997); K. Ingersent, Phys. Rev. B {\bf 54}, 
11936 (1996); C. Gonzalez-Buxton and K. Ingersent, Phys. Rev. B {\bf 57}, 
14254 (1998).
\bibitem{schlot}P. Schlottmann, J. Phys.: Cond. Matter {\bf 10}, 2525 
(1998); P. Schlottmann and A. Zvyagin, Euro. Phys. J. B {\bf 5}, 325 (1998).
\bibitem{Suther}B. Sutherland, Phys. Rev. B {\bf 12}, 3795 (1975); 
P. Schlottmann, Phys. Rev. B {\bf 36}, 5177 (1987).
\bibitem{13}E.K. Sklyanin, J. Phys. A {\bf 21}, 2375 (1988).
\bibitem{14}A. Foerster and M. Karowski, Nucl. Phys. B {\bf 396}, 11 (1993); 
C. Destri and H.J. de Vega, Nucl. Phys. B {\bf 361}, 361 (1992); 
ibid {\bf 374}, 692 (1992).
\bibitem{15}M. Takahashi, Prog. Theor. Phys. {\bf 46}, 401 (1971).
\bibitem{16}N. Andrei, K. Furuya and J.H. Lowenstein, Rev. Mod. Phys. 
{\bf 55}, 331 (1983).
\bibitem{17}H. Johannesson, Phys. Lett. A {\bf 116}, 133 (1986).
\bibitem{18}P. Schlottmann, Phys. Rep. {\bf 181}, 1 (1989).
\end{references}
\end{document}